# Potential of on-demand services for urban travel

Nejc Geržinič[1] · Niels van Oort[1] · Sascha Hoogendoorn-Lanser[2] · Oded Cats[1] ·
Serge Hoogendoorn[1]

## Abstract

On-demand mobility services are promising to revolutionise urban travel, but preliminary studies are showing they may actually increase total vehicle miles travelled, worsening road congestion in cities. In this study, we assess the demand for on-demand mobility services in urban areas, using a stated preference survey, to understand the potential impact of introducing on-demand services on the current modal split. The survey was carried out in the Netherlands and offered respondents a choice between bike, car, public transport and on-demand services. 1,063 valid responses are analysed with a multinomial logit and a latent class choice model. By means of the latter, we uncover four distinctive groups of travellers based on the observed choice behaviour. The majority of the sample, the Sharing-ready cyclists (55%), are avid cyclists and do not see on-demand mobility as an alternative for making urban trips. Two classes, Tech-ready individuals (27%) and Flex-ready individuals (9%) would potentially use on-demand services: the former is fairly time-sensitive and would thus use on-demand service if they were sufficiently fast. The latter is highly cost-sensitive, and would therefore use the service primarily if it is cheap. The fourth class, Flex-sceptic individuals (9%) shows very limited potential for using on-demand services.

## Introduction

A recent societal trend that made its way into the transportation domain is the sharing economy (Cagle 2019). One of its exemplars, present around the world, is the ridesourcing company Uber (also known as a Transportation Network Company or TNC), which began operations in 2009. Since then, a myriad of TNCs have appeared around the world (Lyft, DiDi, Grab etc.)

Recent findings suggest that ridesourcing companies may be a contributing factor to worsening traffic conditions in cities, particularly in downtown areas (Balding et al. 2019;

✉ Nejc Geržinič
n.gerzinic@tudelft.nl

1    Department of Transport & Planning, Delft University of Technology, Delft, Netherlands

2    Delft University of Technology, Delft, Netherlands

Springer

Erhardt et al. 2019; Rodier 2018). One reason for this is a relatively large proportion (30–60%) of empty vehicle miles travelled (VMT) while roaming and waiting for travellers (Henao and Marshall 2018). The other reason lies in ridesourcing substituting other existing modes. While the majority seem to replace car and taxi trips (40–70%), up to 30% of TNC users would have used public transport otherwise while 8–22% would not have travelled at all (Rodier 2018). Henao and Marshall (2018) looked at the impact of mode substitution and found that almost 37% of ridesourcing VMT is generated by users who would have otherwise gone by public transport, active modes (walking and cycling) or not have travelled at all. They conclude that, were it not for TNCs, these travellers would have generated almost 50-times fewer VMT.

While the increasing adoption of ridesourcing services seems to exacerbate congestion in cities, the technology and service they introduce provide opportunities for improving accessibility and equity in urban mobility. One of the possibilities are pooled on-demand trips—as opposed to private trips—where passengers with a similar trajectory and departure time are combined and travel with the same vehicle, increasing vehicle occupancy and reducing VMT (Henao and Marshall 2018). Developing the necessary sharing algorithms and showcasing the benefits is a central topic of many papers in the field (Alonso-Mora et al. 2017; Bischoff et al. 2018; Inturri et al. 2019; Kucharski and Cats 2020; Lokhandwala and Cai 2018; OECD 2015; Ota et al. 2015; Sayarshad and Oliver Gao 2018).

For ridesourcing services to offer an attractive pooling alternative it is essential to get a better understanding of travellers' preferences. Firstly, we need to have good insight into the attitudes towards on-demand mobility on one hand and alternative modes in the urban environment on the other hand. This will allow assessing how the use of certain modes will be affected with the introduction of on-demand mobility services. Secondly, to get as many people as possible to use pooled ridesourcing, we also need to identify travellers' willingness to share rides and how this relates to other travel preferences.

In this paper we use the terms 'ride-hailing' and 'ridesourcing' when referring to services and companies such as Uber, Lyft, DiDi etc. These however are not the only types of on-demand transport services. Some incumbent public transport companies offer demand responsive transit (DRT) or microtransit in low demand areas, offering a flexible alternative to the fixed bus lines, such as Mokumflex (Coutinho et al. 2020) and Breng flex (Alonso-González et al. 2018). From a traveller's perspective, microtransit and (pooled) ride-hailing are largely similar, as in both cases travellers order a ride using a smartphone app, website or by calling, for a trip from A to B without transferring, that the vehicle would not have made otherwise. A notable difference between the two services is that microtransit may not be able to offer a door-to-door service and thus a certain access time is necessary. As we account for access walking time in the study, these transport services are from here on referred to as *on-demand services* or *Flex* services. They can be ordered either as *private* (the rider requests a direct trip, without other passengers) or *pooled* (the vehicle may take small detours to pick up and drop off other passengers along the route) and are referred to as such.

The rest of the paper is structured as follows. The current literature covering on-demand services, their role in the current mobility environment and analyses related to adoption and use of on-demand services are presented in Chapter 2. Chapter 3 describes the methodology used to address the identified gaps in literature. The first subchapter elaborates on the survey design method and the data gathering, and the second subchapter presents the modelling framework used to analyse the gathered data and the third then describes the obtained sample. Chapter 4 then showcases the results of the model estimation, the identified user groups and the potential of the different classes to use Flex services. Finally, the results and policy implications of the emergence of on-demand services are discussed in Chapters 5 and 6.

## Literature review

The impact of on-demand services on the use of public transportation has already been the topic of several studies. Ride-hailing services can complement public transport by offering first/last mile access/egress service, making PT more attractive (Mohamed et al. 2019). Additionally, on-demand services can be used for making trips that are not served well by public transport, like tangential trips carried out by Kutsuplus in Helsinki (Haglund et al. 2019). The Bridj service in Sydney offers both feeder service to train stations during peak times and stop-to-stop services during off-peak times (Perera et al. 2019). Another way on-demand services can complement public transport is to provide services in late evenings and early mornings. Many ride-hailing trips are already made at these times (Fridays and weekends in the late evening), mostly for leisure purposes (King et al. 2020; Mohamed et al. 2019; Tirachini and del Río 2019). Given that these trips happen at times when public transport services are mostly limited or non-existent, some studies suggest that car and taxi trips are the most likely to be affected. On the other hand Young et al. (2020) found that a large part of ride-hailing trips substitute public transport. They analysed ride-hailing trips in Toronto and found that a third of all trips were only up to 15 min faster than public transport and only around 25% were more than half an hour quicker.

Most studies on the matter are inconclusive and assert that ride-hailing is being used as both a substitution for PT and a complementary service to PT (Gehrke et al. 2019; King et al. 2020; Tirachini and del Río 2019). With respect to the impact of ride-hailing on active modes, the results are less clear and also less transferrable to the Dutch context, as most studies on this topic were carried out in areas where active modes are less prominent. Notably, Gehrke et al. (2019) report that in Boston, active modes are mainly being replaced by on-demand services in bad weather, whereas in good weather conditions, ride-hailing was found to mainly replace car and public transport trips, indicating that active modes are not severely impacted. This finding is also corroborated by Henao and Marshall (2018) for data from Denver, reporting that about 12% of ride-hailing trips are substituting active modes. Neither study however reports the modal share of active modes in their respective cities, so the scale of impact on active modes is unclear.

To better understand the behavioural trade-off travellers make when faced with an option to choose for on-demand services, several studies employed stated preference (SP) experiments. Most of these studies applied either a multinomial logit (MNL) a mixed logit (ML) model formulation (Table 1). They mostly introduce pooled on-demand next to the car and/or public transportation. Liu et al. (2018) carried out a study in New York, comparing pooled and private ride-hailing with car and public transport and found the highest preference for car, followed by PT, with on-demand services at the bottom (private being preferred over pooled). The same preference order amongst modes was also found in studies in Chicago (Frei et al. 2017), Ann Arbor, Michigan (Yan et al. 2019) and in North England (Ryley et al. 2014), while Choudhury et al. (2018), who carried out the survey in Lisbon, found a higher preference for public transport compared to car, while on-demand services were still less preferred. They hypothesize that the former can be attributed to the high quality public transport and congested roads in the city centre. They compared a much larger number of alternatives and trip contexts by employing a novel four-step

**Table 1** Overview of studies using stated preference data collection for analysing on-demand mobility

| | Data collection | Choice model estimated | Geographical location | Pooled on-demand | Private on-demand | Cycling | Public transport | Car |
|---|---|---|---|---|---|---|---|---|
| Frei et al. (2017) | SP | MNL and ML | Chicago | X | | | X | X |
| Liu et al. (2018) | SP | ML | New York | X | X | | X | X |
| Yan et al. (2019) | RP and SP | MNL and ML | Ann Arbor (Michigan) | X (*SP only*) | | X | X (*RP only*) | X |
| Ryley et al. (2014) | SP | ML | North England | X | | | X[a] | X[a] |
| Choudhury et al. (2018) | SP | ML (and nested logit) | Lisbon | X | X | | X | X |
| Alonso-González, et al. (2020a, b, c) | SP | LCCM | The Netherlands | X | X | | | |
| This study | SP | LCCM | The Netherlands | X | X | X | X | X |

[a] In their SP survey, Ryley et al. (2014) showed respondents one existing mode (public transport or car) alongside DRT

multi-dimensional pivot SP experiment. Alongside the previously mentioned modes, they also considered car rental, bus, express bus, train/metro, bus + train/metro and a park-and-ride alternative. Their findings show a higher willingness-to-pay for car-based modes (including on-demand mobility) and lower for public transport modes. A similar Willingness to Pay (WtP) pattern was observed by Frei et al. (2017), who also modelled the impact of weather on the attractiveness of ride-hailing and found that colder and rainy weather makes it more attractive for respondents, compared to using the car or public transport.

To the best of our knowledge, Yan et al. (2019) are the only ones to carry out an SP survey of an on-demand service, where cycling was also offered as an alternative. They jointly modelled the revealed preference (RP) behaviour and SP survey responses of the university faculty members and found a relatively high preference for on-demand service in the SP only analysis (above car and cycling) and a somewhat low preference, using combined RP-SP data, where on-demand is behind the car and transit, while still preferred over cycling. A major limitation of their study is that on-demand services were offered for free and the only cost parameter in the survey was car parking cost, hampering a monetary evaluation in the form of willingness to pay.

Another avenue of exploring the attractiveness and adoption potential of on-demand services is through different latent class clustering methods. Applying a latent class adoption model (Alemi et al. 2019) on the California Millennials dataset (Circella et al. 2016), to analyse the potential of the respondents to adopt ride-hailing revealed that among the three uncovered classes, the most likely to adopt ride-hailing are those who are highly educated and independent Millennials, living in urban areas and are without children, whereas the least likely to adopt it are those living in rural areas.

Pooling on-demand trips is also associated with both physically sharing the vehicle with other people and can lead to an uncertain increase in both waiting and travel time. The Dutch population seems to be fairly evenly split between those more open to sharing (usually to reduce their trip cost) and those who are not, either for privacy reasons or time-related reasons (Alonso-González et al. 2020a, b, c). With respect to the valuation of travel time variability, Alonso-González et al. (2020a, b, c) report that a large majority of the population has a balanced time–cost sensitivity next to two smaller groups that are either more time-sensitive or more cost-sensitive. The perception of reliability was found fairly similar across the classes, with the variable travel and waiting times being valued between 0.5 and 1-times the value of actual travel and waiting time.

This study adds to the existing literature by analysing respondents' mode choice in a setting where both private and pooled on-demand services exist alongside cycling, cars and public transport (Table 1). In contrast to most literature undertaking such research, heterogeneity among decision-makers is accounted for by employing a latent class choice model (LCCM), instead of a mixed logit model specification. Using an LCCM enables the identification of market segments and through performing a posterior analysis, we obtain detailed insights into the different classes' socio-demographic characteristics, their attitudes and current travel behaviour.

## Methodology

To analyse the role of on-demand mobility services and the perceptions and preferences related to them, we employ a stated preference approach. On-demand services are not yet widespread in the Netherlands, and while many people have heard of such services, many have not used such services yet, as highlighted by Bronsvoort et al. (2021). Given this circumstance, we opt for an SP choice experiment.

## Survey design

To elicit respondents' decision-making behaviour with respect to on-demand mobility, a stated preference survey is constructed and conducted. In the survey, on-demand mobility is labelled as "*Flex*" to ease communication (Alonso-González et al. 2020c), as such services typically have the "Flex" suffix (*Brengflex* (Arnhem-Nijmegen), *TwentsFlex* (Rijssen-Holten), *Mokumflex* (Amsterdam) etc.). Flex is positioned next to the most commonly used modes for urban trips in the Netherlands, namely the bike, car and public transport. Walking was excluded due to the longer trip distance and also to reduce the number of alternatives shown. Modes are compared based on in-vehicle time, walking time, waiting time, trip cost and whether or not the ride is shared. In the Netherlands more than half of all the trips are up to five kilometres long and another 20% are between 5 and 15 kms (de Graaf 2015). To determine the attribute levels, a trip that is approximately five kilometres long is selected and a range of values for a trip of such distance is obtained from Google (n.d.). All but one attribute ("*Type of ride*") are described with three levels, to allow the analysis of non-linear attribute evaluation. The four modes along with their respective attributes and levels can be seen in Table 2. This design is used to assess preferences for two different trip purposes: a commute trip (travelling to work or education) and a leisure trip (travelling for recreation, visiting friends / family). Each choice set contains five alternatives, with two "*Flex*" alternatives alongside the three existing modes. This, together with the "*Type of ride*" attribute allows a direct comparison between two "*Flex*" alternatives, be it two shared options, two private options or one shared and one private (as seen in Fig. 1).

To avoid making a-priori assumptions on parameter values and willingness-to-pay (WtP), an orthogonal design is selected for this survey (Walker et al. 2018). Findings from studies on the perceptions of on-demand mobility are not fully in agreement, and if the actual preferences among the respondents do not align with those of the priors, a D-efficient design might become highly inefficient (Walker et al. 2018). An orthogonal design is generated in the software tool Ngene (ChoiceMetrics 2018). The design is made up of 72 choice sets, which are blocked over 12 blocks of six choice sets. Each respondents is randomly allocated to two different blocks, one for a commute and one for a leisure trip purpose. The 12 blocks are randomly allocated to respondents to guarantee an approximately equal number of observations per block and trip purpose. One example choice set (originally in Dutch) can be seen in Fig. 1. Respondents who do not have access to a car in their household are presented with choice sets without the car alternative. It should be noted that the "*Waiting time*" attribute for public transport and "*Flex*" are not presented to respondents in the same way for the two modes. For public transport, waiting is said to be endured at a stop, while for the on-demand service, respondents are given a summary of their trip, indicating how many minutes they need to leave their origin (home) in and then possibly walk to a pick-up point. With that in mind, the two different waiting times are shown to respondents in distinctly different ways. For transit, the word "*waiting*" (*wachten* in Dutch) is used and positioned between "*walking*" and "*in-vehicle time*" (*lopen* and *OV tijd* in Dutch respectively), whereas for "*Flex*", the attribute is shown first as "*depart in*" (translated from the Dutch *vertrek over*).

**Table 2** Modes, attributes and attribute levels used in the Urban survey

| Attributes | Bike | Public transport | Car | Flex |
|---|---|---|---|---|
| Walking time (min) | – | 1, 5, 9 | 0, 5, 10 | 0, 3, 6 |
| Waiting time (min) | – | 1, 5, 9 | – | 1, 5, 9 |
| In-vehicle time (min) | 12, 16, 20 | 8, 12, 16 | 8, 12, 16 | 8, 12, 16 |
| Type of ride | – | – | – | Shared, private |
| Cost (€) | – | 0.5, 2, 3.5 | 1, 5, 9 | 2, 5, 8 |

**Fig. 1** Example of the choice set shown to respondents (translated to English)

In addition to the twelve choice tasks, respondents are presented with 16 attitudinal statements (Table 3). The goal of the statements is to understand respondents' preparedness to use on-demand mobility services. In investigating the drivers of MaaS and Flex adoption, Alonso-González, et al. (2020a, b, c) used three groups of attitudinal statements, categorized by Durand et al. (2018) into: (1) Mobility integration, (2) Shared mobility modes and (3) Mobile applications. While the goal of this research is to better understand Flex-readiness, a similar setup of attitudinal statements is used, as there are many similarities in the willingness to use of MaaS and on-demand mobility. The formulated attitudinal statements fall into one of four categories:

1. Use of apps
2. Mobility integration
3. Sharing a ride
4. Sharing economy

The first category investigates the understanding and willingness to use smartphone based travel planning applications and certain app features, like making purchases and GPS navigation. Mobility integration, similar to the study by Alonso-González, et al. (2020a, b, c), considers the attitude towards multimodal travel and public transport, as well as the attitude towards not having to drive a car and if that is seen as beneficial or not. Integration with other modes is relevant, as an on-demand service often provides first/last mile connectivity. Previous studies have also found that users of on-demand services tend to

**Table 3** Attitudinal statements on pooling rides, travel planning and the sharing economy

| Category | | Statement |
|---|---|---|
| Use of (travel planning) apps | 1 | I find it difficult to use travel planning apps[a] |
| | 2 | Using travel planning apps makes my travel more efficient[a] |
| | 3 | I am willing to pay for transport related services within apps |
| | 4 | I do not like using GPS services in apps because I am concerned for my privacy |
| Mobility integration | 5 | I am confident when travelling with multiple modes and multiple transfers |
| | 6 | I do not mind infrequent public transport, if it is reliable |
| | 7 | I do not mind having a longer travel time if I can use my travel time productively[b] |
| | 8 | Not having to drive allows me to do other things in my travel time[b] |
| Sharing a ride | 9 | I am willing to share a ride with strangers ONLY if I can pay a lower price[b] |
| | 10 | I feel uncomfortable sitting close to strangers[b] |
| | 11 | I see reserving a ride as negative, because I cannot travel spontaneously |
| Sharing economy | 12 | I believe the sharing economy is beneficial for me |
| | 13 | I believe the sharing economy is beneficial for society |
| | 14 | Because of the sharing economy, I use traditional alternatives (taxis, public transport, hotels,…) less often |
| | 15 | Because of the sharing economy, I think more carefully when buying items that can be rented through online platforms |
| | 16 | I think the sharing economy involves controversial business practices (AirBnB renting, Uber drivers' rights,…) |

[a]Adapted from (Lu et al., 2015)

[b]Adapted from (Lavieri & Bhat, 2019)

The remaining statements were formulated for the purpose of this study be more open to using a variety of transport modes and more often travel long-distance (Alemi et al. 2018). Statements on "Sharing a ride" are meant to capture the social aspect to sitting (close) to strangers. It is also used for a direct comparison to the stated choice survey outcome on the discount required for users to opt for pooled services. Finally, five statements pertaining to the sharing economy are included. As many on-demand mobility services are amongst the most well-known examples of the sharing economy (Uber, Lyft and DiDi being the most prominent), we want to see if there are any links to be made with respect to respondents' attitudes towards the sharing economy and their travel behaviour. Each statement is evaluated on a 5-point Likert scale, and an additional 'No opinion' option is added to each statement. The 'Neutral' and 'No opinion' attitudes are coded as 0, 'Agree' and 'Fully agree' as 1 and 2 respectively and 'Disagree' and 'Fully disagree' as −1 and −2 respectively.

Familiarity with six different shared mobility and sharing economy services is also inquired, with the services and the examples given, shown in Table 4. Respondents are asked to indicate their familiarity with the service on a 5-point scale:

1. Never heard of it
2. Familiar with it, but never used it
3. Used it once
4. Used it a few times
5. Use it regularly

As the data is obtained through the Dutch Mobility Panel (MPN—Mobiliteitspanel Nederland), there is no need to ask respondents for basic socio-demographic information, as this is regularly updated for the panel members (Hoogendoorn-Lanser et al. 2015).

## Model estimation

The model is estimated with Biogeme, an open source Python package (Bierlaire 2020). To analyse respondents' travel preferences, two types of discrete choice models are employed. Firstly, to get an overview of choice behaviour, understand the preferences towards different modes and potential non-linear perceptions of attributes, several different MNL models are estimated. Secondly, to analyse different potential user groups and their respective attitudes and preferences, we perform a market segmentation by means of estimating a latent class choice model (LCCM). All models are estimated under the assumption that users select the alternative with the goal of maximising the utility of their choice (McFadden 1974).

**Table 4** Services for which respondents were asked to indicate their familiarity (including the presented examples)

|   | Type of (sharing economy) service | Examples shown |
|---|---|---|
| 1 | How familiar are you with car sharing? | Snappcar, Greenwheels, car2go |
| 2 | How familiar are you with bike / scooter sharing? | Mobike, OV fiets, Felyx |
| 3 | How familiar are you with flexible public transport? | Twentsflex, Bravoflex, U-flex, Delfthopper |
| 4 | How familiar are you with ride-hailing? | Uber, ViaVan |
| 5 | How familiar are you with food delivery services? | Thuisbezorgd, Deliveroo, Foodora, UberEATS |
| 6 | How familiar are you with home rental services? | AirBnB, HomeStay, Couchsurfing |

A generic-parameter model (Eq. 1) is estimated as a benchmark for the more detailed models that follow. All attributes ($a$) are coded by a single generic parameter ($\beta_a$) for the four different modes ($m$), with the exception of waiting time, which is modelled separately for public transport and Flex (see the Survey design section above for the argumentation). The ASP model (Eq. 2) expands on the GP model by splitting the parameters for each individual mode ($\beta_{m,a}$), allowing a comparison of how the same attribute is perceived across different modes. The DCP model (Eq. 3) is estimated to uncover and analyse potential non-linarites in the perceptions of attributes. This is done by modelling the utility contribution of each attribute level ($X_{m,a}$) by means of its own parameter ($\beta_{Xm,a}$), with one of the levels being fixed to 0. Based on the outcomes of the three models, other model specifications are tested to obtain the most parsimonious model.

*Equation 1: GP model formulation*

$$V_m = ASC_m + \sum\nolimits_a \beta_a \bullet X_{m,a} \tag{1}$$

*Equation 2: ASP model formulation*

$$V_m = ASC_m + \sum\nolimits_a \beta_{m,a} \bullet X_{m,a} \tag{2}$$

*Equation 3: DCP model formulation*

$$V_m = ASC_m + \sum\nolimits_a \beta_{X_{m,a}} \bullet X_{m,a} \tag{3}$$

To analyse respondent heterogeneity, a latent class choice model is chosen (Greene and Hensher 2003). Unlike a mixed logit model, a latent class model enables for the estimation of individual parameters for each obtained class, resulting in a straightforward interpretation of the classes and a clear distinction between them. The added value of a latent class model is also the possibility of making a posterior analysis of the attitudinal and socio-demographic characteristics for each of the classes. The model formulation of the LC model is presented in Eq. 4. The probability of respondent $n$ to select alternative $i$ is obtained by summing the probability of this alternative being selected, given the different parameter estimates ($\beta$) in the different latent classes ($s$). The class specific choice probabilities ($P_n(i|\beta)$) are multiplied with the class allocation probability ($\pi_{ns}$). In its simplest form, the class membership function is static (Hess et al. 2008), meaning that only a constant ($\delta_s$) is used. Additionally, the class membership function of the latent class model may include socio-demographic data as a predictor for allocating individuals to different segments of the population. In this study, the goal is to group individuals based purely on their stated preferences, so that travel behaviour preference heterogeneity of individuals within a group is as small as possible, while the difference between groups is as large as possible. A static class membership function is employed in this study. For the individual class formulations, the model which proves as most parsimonious among the MNL model formulations is used in the LC model, to allow for faster estimation, ease of interpretation and to guarantee that sufficient parameters of interest can be identified.

*Equation 4: Formulation of the LC model*

$$P_n(i|\beta) = \sum\nolimits_{s=1}^{S} \pi_{ns} \bullet P_n(i|\beta_s)$$

*Equation 5: Formulation of the class allocation probability*

$$\pi_{ns} = \frac{e^{\delta_s}}{\sum_{l=1}^{S} e^{\delta_l}}$$

For the 16 attitudinal statements, an exploratory factor analysis (EFA) is performed, to analyse correlations between the responses to statements, to simplify the interpretation of said responses and also to simplify the interpretation of differences between the latent classes with respect to their attitudes towards ride-sharing and the sharing economy. To perform the EFA, the "factor_analyzer" package for Python is used (Briggs 2019).

To obtain socio-demographic information for the different latent classes, a posterior probability analysis is carried out. Individuals are probabilistically allocated to each of the classes, based on how well the class-specific parameters capture the respondent's observed choices. Based on this probability, the socio-demographic, attitudinal and travel behaviour characteristics are aggregated per class.

## Data collection

The survey was administered to the participants of the Netherlands Mobility Panel (MPN) (Hoogendoorn-Lanser et al. 2015), between February 10th and March 1st in 2020. In total, 1,200 respondents took part in the survey, which was reduced to 1,063 after processing and cleaning the raw data. This was done by removing:

1. Responses with incomplete choice tasks
2. Responses that were completed within fewer than three minutes
3. Respondents with the same answer to all attitudinal questions

The socio-demographics of the sample and the Dutch population are shown in Table 5 (Centraal Bureau voor de Statistiek 2020). The sample is largely representative of the Dutch population. We do note a slight overrepresentation of older individuals in the survey compared to the population. Household income differs as well, which can partly be explained by respondents having the option to not disclose their income in the survey, while the census data has a complete overview of everyone's incomes.

With respect to COVID-19, the first patient in the Netherlands was diagnosed on the 27th of February (Rijksinstituut voor Volksgezondheid en Milieu (RIVM) 2020) and the first lockdown measures announced on March 12th (NOS 2020). We therefore believe that it is unlikely that the epidemic influenced the decision-making of the respondents.

## Results

Outcomes of the four different MNL model specifications, as well as the latent class model, are shown in Table 6. The 4-class latent class model significantly outperformed the other three models. Of the MNL models, the DCP model achieves the highest model fit and adjusted rho-squared value. Surprisingly, the ASP model performs relatively well compared to the DCP model, with a final log-likelihood only 13 points lower and essentially no difference in the value of the adjusted rho-squared. Performing a likelihood ratio test, the DCP model is found to be superior to the ASP model. Considering the BIC value on the other hand, the ASP model is superior, as it achieves a similarly high model fit with fewer parameters. Testing different model specifications resulted in an additional model (labelled

**Table 5** Comparison of socio-demographic variables for the survey sample and the Dutch population

| Variable | Level | Sample (%) | Population (%) |
|---|---|---|---|
| Gender | Female | 52 | 50 |
| | Male | 48 | 50 |
| Age | 18–34 | 21 | 27 |
| | 35–49 | 20 | 23 |
| | 50–64 | 30 | 26 |
| | 65+ | 29 | 24 |
| Education[a] | Low | 30 | 32 |
| | Middle | 41 | 37 |
| | High | 29 | 31 |
| Urbanisation level | Very highly urban | 23 | 24 |
| | Highly urban | 32 | 25 |
| | Moderately urban | 17 | 17 |
| | Low urban | 20 | 17 |
| | Not urban | 8 | 17 |
| Household income[b] | Below average | 24 | 26 |
| | Average | 50 | 47 |
| | Above average | 12 | 27 |
| | Unknown | 14 | 0 |
| Employment status | Working | 50 | 51 |
| | Not working | 50 | 49 |
| Household size | One person | 22 | 17 |
| | 2 or more | 78 | 83 |

Source for the population data: (Centraal Bureau voor de Statistiek, 2020)

[a]Low: no education, elementary education or incomplete secondary education

Middle: complete secondary education and vocational education

High: bachelor's or master's degree from a research university or university of applied sciences

[b]Below average: below modal income ($< €29,500$)

Average: 1–2×modal income (€29,500–€73,000)

Above average: Above 2×modal income ($> €73,000$)

**Table 6** Outcomes of models with different parameter specifications

| | GP model | LC-base model | ASP model | DCP model | Latent class model |
|---|---|---|---|---|---|
| Number of estimated parameters | 10 | 11 | 19 | 31 | 47 |
| Final log-likelihood | − 11,595.91 | − 11,568.26 | − 11,443.90 | − 11,430.83 | − 6,653.10 |
| Adjusted Rho-squared | 0.4201 | 0.4220 | 0.4272 | 0.4273 | 0.6652 |
| BIC value | 23,286.35 | 23,240.52 | 23,067.42 | 23,154.72 | 13,633.73 |

the LC-base model), which is also reported in Table 6. It improves the model fit of the GP model by 28 LL-points with a single parameter. The improvement from the LC-base to the ASP model is 124 LL-points, utilising eight additional parameters. While the BIC and LRT both prove the ASP model is superior to the LC-base model, the marginal contribution of the additional estimated parameters is less than the LC-base model achieves. Given these outcomes, the following section will focus primarily on the interpretation of the LC-base model results. The respondents' familiarity with shared services and their replies to the 16 statements are also discussed. In Sect. 4.2, the latent class model is then analysed and each of the four classes is discussed on their respective taste parameters, attitudinal statements, socio-demographic characteristics and current travel behaviour.

## Results of discrete choice model

The parameter estimates of the LC-base model are presented in Table 7, with the outcomes of the ASP and DCP models shown in Appendix A in Tables 11 and 12. With the exception of the Flex waiting time parameter, all other are highly significant. This parameter turns out insignificant in all estimated MNL models, while in the DCP model, a waiting time of 5-min seems to be perceived more negatively than a 9-min waiting time. One possible explanation for this is in the way Flex waiting time was specified in the survey: waiting at home, which can be seen as hidden waiting time. A 5-min duration can be seen as period of time that cannot really be spent on doing anything—just waiting—while nine minutes could already be enough time to accomplish a quick errand at home, meaning the 'waiting' time is well spent and therefore does not have such a high disutility. A 1-min waiting time is still most preferred, meaning that respondents still preferred the shortest possible waiting time.

In the ASP model, the in-vehicle time parameters for all motorised modes (car, public transport, Flex) are insignificant. This could be due to the relatively small variation in the in-vehicle time attribute levels (8, 12 or 16 min), indicating that respondents do not care about the travel time when the differences are relatively small and only the mode which is used appears to be important. In the LC-base model, the in-vehicle

**Table 7** Model estimation results of the LC-base model

|  | Parameter estimate | Robust t-stat | Significance |
|---|---|---|---|
| Constant [bike] | 0 [ fixed] | | |
| Constant [car] | − 1.216 | − 9.56 | *** |
| Constant [Flex] | − 3.172 | − 21.66 | *** |
| Constant [PT] | − 2.303 | − 17.33 | *** |
| Cost | − 0.148 | − 17.86 | *** |
| In-vehicle time [bike] | − 0.070 | − 11.15 | *** |
| In-vehicle time [other] | − 0.011 | − 2.02 | ** |
| Walking time | − 0.047 | − 9.59 | *** |
| Waiting time [Flex] | − 0.014 | − 1.24 | |
| Waiting time [PT] | − 0.039 | − 4.22 | *** |
| Sharing [Flex] | − 0.215 | − 2.82 | *** |
| Leisure trip * cost | − 0.022 | − 2.53 | ** |

***$p \leq 0.01$, **$p \leq 0.05$, *$p \leq 0.1$

time for all three motorised modes is combined into a single parameter, distinguishing it only from the cycling time parameter. Distinguishing two in-vehicle time parameters (cycling and motorised modes) is also the only difference between the GP model and LC-base model, showing that estimating a separate cycling time parameter significantly improves model fit. This is in part due to the fact that the cycling alternative has no other attributes apart from travel time. Cycling time is thus perceived far more negatively than in-vehicle time in motorised modes: six times more.

The mode specific constants capture all other factors not included in the modelled attributes, which are associated with a specific mode, with all other attributes being equal. Table 7 reveals that the ASC for bike (fixed to 0) is the highest, followed by car, PT and the Flex ASC having the lowest value. This is mostly in-line with findings reported in the literature (Frei et al. 2017; Liu et al. 2018), although most did not include cycling and the one that did (Yan et al. 2019) found cycling to have the lowest ASC. This likely has to do with the survey being set in Michigan, USA, where the cycling conditions and culture are very different from those in the Netherlands. Choudhury et al. (2018) found car to have a lower ASC than the PT alternatives, but they also state that this is due to the survey being conducted in Lisbon.

Comparing the ratios of different travel time components, walking time is perceived four times more negatively than in-vehicle time, whereas waiting time for PT is seen as more than three times as negative. This is higher than expected and higher than reported in research (Wardman 2004), but could be a consequence of the relatively low disutility associated with in-vehicle time, partly due to the limited differences in travel time attribute levels.

Of prime interest for this research is the willingness-to-share on-demand services. Sharing is explored both as a dummy variable (as seen here) and by interacting it with in-vehicle time, as a perceived in-vehicle time multiplier. A superior model fit is achieved when the former model specification is used. From the parameter ratios in the case of the alternative-specific model, we can see that respondents are willing to pay up to €1.45 (*Sharing [FLEX] / Cost*) more for a private trip, which is higher than the 0.41€ reported by Alonso-González, et al. (2020a, b, c) for respondents' willingness to pay to avoid sharing with one or two other people, but less than over $6 reported by Liu et al. (2018). The former study analysed sharing in more detail (different parameters for different numbers of co-riders) and the penalty for only up to two additional passengers is likely less than for a full vehicle. The difference from the latter study could be due to the difference in cultural context (survey carried out in the USA, as opposed to the Netherlands for this study). Respondents are also willing to walk almost 5 min farther for a private ride (*Sharing [FLEX] / Walking time*), or travel up to 20 min longer in a private ride as opposed to a shared ride (*Sharing [FLEX] / In-vehicle time [other]*).

Several different model specifications are also explored with respect to trip purpose (commute and leisure). The trip purpose is interacted with all estimated parameters separately. The best model fit and a highly significant parameter estimate is obtained when the trip purpose is interacted with trip cost. In line with other findings (Alonso-González et al. 2020c; Choudhury et al. 2018), respondents are more cost-sensitive in leisure trips than in their commute trips. Respondents seem to be approximately 15% more cost sensitive when making leisure trips as opposed to commute trips, meaning they are willing to pay less when travelling for leisure.

## Service familiarity and exploratory factor analysis

As shown in Fig. 2, respondents are mostly familiar with sharing economy and shared transport services. Flexible public transport services on the other hand are much less well known than any of the other services, with 51% of the respondents never having heard of it and only 2% using it at least once. Ride-hailing and bike sharing services are most familiar to respondents, although still only around 10% of them have ever used it. Food delivery is the most used sharing economy service, with over 40% having used it at least once before.

The 16 attitudinal statements (Fig. 3) are used to elicit respondents' readiness to use shared on-demand services. In general, respondent agree that apps are easy to use and make travel more efficient, they prefer not making in-app purchases. Participants overall agree that not having to drive gives opportunities to better spend one's travel time and are also willing to travel longer, if that means they can better use their travel time. Respondents are largely in agreement that they are willing to use a shared service only if they receive a discount (statement 9). There is also a clear indication that they feel uncomfortable sitting close to strangers (statement 10). Attitudes towards the sharing economy reveal that respondents are quite optimistic for what the sharing economy has to offer to society, but for the most part, do not see many direct benefits for themselves. They do however, believe that in some cases, the sharing economy can lead to controversial business practices.

For an easier interpretation of the attitudinal statements, an exploratory factor analysis is performed. A KMO score of 0.78 is obtained, indicating that an EFA can indeed be performed (Ledesma et al. 2021). As mentioned in Sect. 3.1, the statements had 6 possible answers, including a 5-point Likert scale and a "No opinion" option. The latter present a potential issue for performing the EFA. It is decided that for the sake of the EFA, "No opinion" answers are converted to the "Neutral" opinion in the 5-point Likert scale. While this is not ideal, and we certainly cannot state that they represent the same sentiment of the respondents, removing all replies with a "No opinion" answer results in a loss of almost 40% of observations (from 1063 to 660). It also severely impacts the interpretation of the outcomes.

To determine the number of factors to estimate, a scree plot is generated. From the plot five factors seem to be the optimal number for the analysis. However, one of the five factors has only one statement with a strong load. Because of that, a 4-factor EFA is seen as preferred. The factor loadings for the four factors are presented in Fig. 4. The four factors largely reflect the four topics which the statements were formulated to capture. S16, unlike the other four sharing-economy-related statements does not seem to

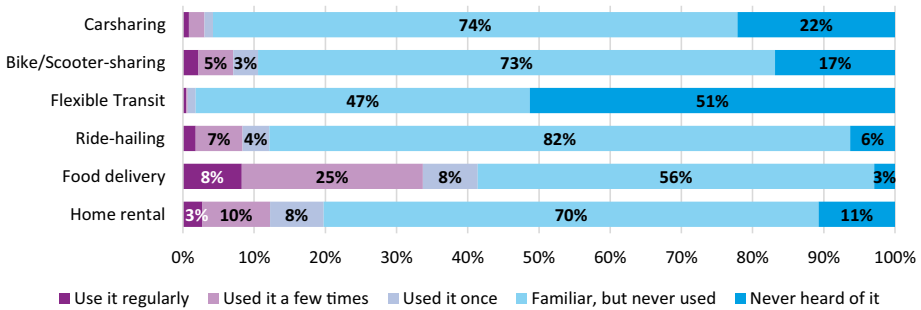

**Fig. 2** Familiarity and frequency of use w.r.t. different shared transport and sharing economy services

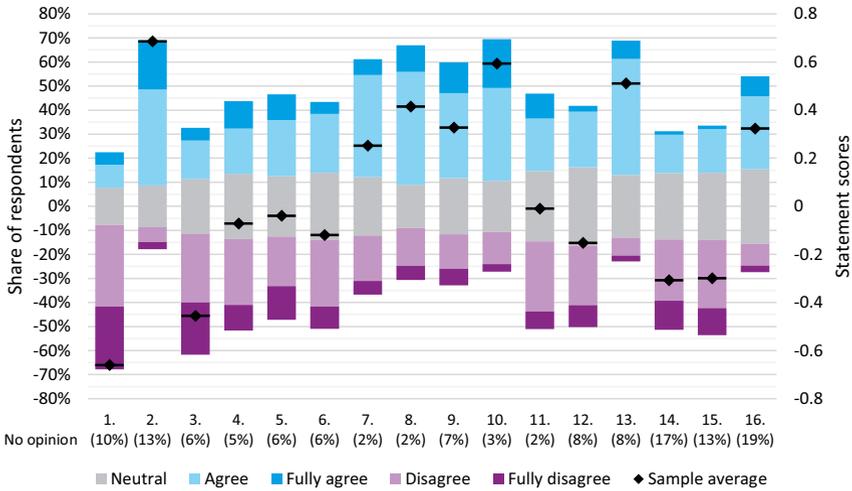

**Fig. 3** Responses to the attitudinal statements, including the average score for each of the statements

1. I find it difficult to use travel planning apps.
2. Using travel planning apps makes my travel more efficient.
3. I am willing to pay for transport related services within apps.
4. I do not like using GPS services in apps because I am concerned for my privacy.
5. I am confident when travelling with multiple modes and multiple transfers.
6. I do not mind infrequent public transport, if it is reliable.
7. I do not mind having a longer travel time if I can use my travel time productively.
8. Not having to drive allows me to do other things in my travel time.
9. I am willing to share a ride with strangers ONLY if I can pay a lower price.
10. I feel uncomfortable sitting close to strangers.
11. I see reserving a ride as negative, because I cannot travel spontaneously.
12. I believe the sharing economy is beneficial for me.
13. I believe the sharing economy is beneficial for society.
14. Because of the sharing economy, I use traditional alternatives (taxis, public transport, hotels,…) less often.
15. Because of the sharing economy, I think more carefully when buying items that can be rented through online platforms.
16. I think the sharing economy involves controversial business practices (AirBnB renting, Uber drivers' rights,…).

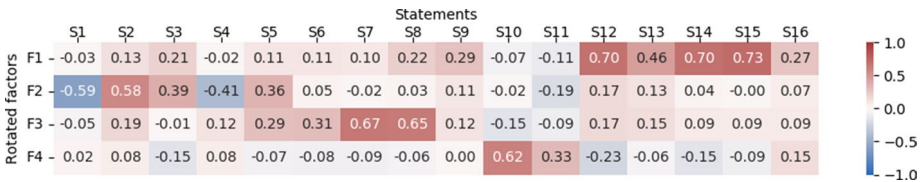

**Fig. 4** Factor loadings from an exploratory factor analysis on the 16 attitudinal statements

load strongly onto any factor. Similarly, S6 and S9 are not very strongly associated with the other statements pertaining to»Mobility integration« and»Sharing a ride« respectively. Finally, it is interesting to notice that S5 loads stronger onto F2, along with the app-related statements, as opposed to F3, with the other mobility-related statements. Given the statements and factor loadings, the four factors have been given the following names:

**F1: Sharing economy support**

Statements 12–15 asked respondents to consider the added value the sharing economy has for them and soceity, and if they do in fact use traditional services and products less often because of the rise of the sharing economy

**F2: App savvy**

Statements 1–4 all pertain to the use of apps, the difficulty respondents find in using them and the usefulness of apps that the respondents belive they offer

**F3: Travel time use**

Statements 7 and 8 both discuss the duration of travel time, putting forward to respondents the notion of having a potentially longer travel time in exchange for a more effective use of that travel time.

**F4: Dial-a-ride opposition**

Statements 10 and 11 deal with sitting in close proximity of strangers and having to pre-book a ride, both of which can be associated with shared on-demand services, but more specifically dial-a-ride services, as the former do not necessarily have prebooking requirements.

## Results of the latent class model

The model outcomes of a 4-class latent class model are shown alongside the different MNL models in Table 6. In total, models with between two and six classes are estimated. A 6-class latent class model is rejected because two of the classes are too small to serve as a meaningful representation of a market segment (both accounting for only 4% of the population). A 5-class model is rejected because of the very limited interpretability of the classes' characteristics. Models with 2 and 3 classes produced valid and interpretable results, but a model with 4-classes provides insights into a larger number of user groups, while maintaining the interpretability of the results. A latent class model with four classes is therefore chosen for both numerical and interpretation reasons. The MNL model used in the estimation is the LC-base model. The parameter estimates of the four latent classes, along with the class sizes, are presented in Table 8. The classes were given names based on their choice behaviour characteristics and the outcomes of the posterior analysis of the EFA for the 16 attitudinal statements:

- Sharing-ready cyclists (55%)
- Tech-ready individuals (27%)
- Flex-sceptic individuals (9%)
- Flex-ready individuals (9%)

The classes are compared on their parameter estimates and the corresponding willingness-to-pay (WTP) in the following paragraph in Table 9. Each of the four classes is then presented in more detail in the following subchapters. Their attitudinal, travel behaviour and socio-demographic information is discussed and presented in Figs. 5, 6 and Table 10 respectively. These were obtained using the posterior probability analysis.

The '*Sharing-ready cyclists*' seem to be the most sensitive to out-of-vehicle time components, but at the same time the in-vehicle time for motorised modes is found insignificant for them. They also have a fairly strong preference for the bicycle over other modes, with all ASCs being highly significant. The highest WTP for in-vehicle time can be observed for the '*Flex-sceptic individuals*', aligning with the high value

**Table 8** Model estimation results of a 4-class latent class model

| Class size | Latent class 1 | | Latent class 2 | | Latent class 3 | | Latent class 4 | |
| | Sharing-ready cyclists | | Tech-ready individuals | | Flex-sceptic individuals | | Flex-ready individuals | |
| | 55% | | 27% | | 9% | | 9% | |
| | Value | Robust t-stat | Value | Robust t-stat | Value | Robust t-stat | Value | Robust t-stat |
|---|---|---|---|---|---|---|---|---|
| $\delta_s$ | 1.76 | 15.95*** | 1.07 | 8.84*** | 0 | Fixed | − 0.04 | − 0.23 |
| Constant [car] | − 5.659 | − 5.04*** | − 2.094 | − 7.02*** | 11.343 | 3.85*** | 2.937 | 2.83*** |
| Constant [Flex] | − 8.412 | − 6.22*** | − 4.279 | − 12.12*** | − 0.021 | − 0.01 | 1.806 | 1.68* |
| Constant [PT] | − 6.330 | − 5.65*** | − 3.310 | − 10.80*** | 3.063 | 1.45 | 1.349 | 1.28 |
| Cost | − 0.243 | − 3.38*** | − 0.217 | − 8.85*** | − 0.326 | − 2.33** | − 0.424 | − 8.52*** |
| In-vehicle time [bike] | − 0.179 | − 3.76*** | − 0.221 | − 12.03*** | − 0.260 | − 2.50** | − 0.172 | − 2.26** |
| In-vehicle time [other] | − 0.016 | − 0.43 | − 0.040 | − 4.12*** | − 0.084 | − 1.63 | − 0.037 | − 2.57** |
| Walking time | − 0.152 | − 4.68*** | − 0.088 | − 8.61*** | − 0.100 | − 1.50 | − 0.092 | − 5.15*** |
| Waiting time [Flex] | 0.095 | 0.90 | − 0.024 | − 1.06 | 0.025 | 0.41 | − 0.021 | − 1.17 |
| Waiting time [PT] | − 0.111 | − 1.87* | − 0.061 | − 3.87*** | 0.016 | 0.27 | − 0.017 | − 0.92 |
| Sharing [Flex] | − 0.154 | − 0.24 | − 0.234 | − 1.65* | − 0.207 | − 0.37 | − 0.223 | − 1.64 |
| Leisure trip * cost | − 0.099 | − 0.99 | − 0.068 | − 2.30** | − 0.307 | − 4.63*** | 0.005 | 0.13 |

*** $p \leq 0.01$, ** $p \leq 0.05$, * $p \leq 0.1$

**Table 9** Willingness-to-pay for travel time components and ratio of cost sensitivity for different trip purposes

|  | Sharing-ready cyclists | Tech-ready individuals | Flex-sceptic individuals | Flex-ready individuals |
|---|---|---|---|---|
| In-vehicle time (€/h) |  | € 11.15 | € 15.46 | € 5.24 |
| Walking time (€/h) | € 37.48 | € 24.39 |  | € 13.09 |
| PT wait time (€/h) | € 27.42 | € 16.79 |  |  |
| In-vehicle time ratio [cycling/motorised] |  | 5.48 | 3.10 | 4.66 |
| Trip purpose cost ratio [leisure/commute] |  | 1.31 | 1.94 |  |

Parameter ratios are only shown if both parameters are significant at the 90% level of the ASC for car. They do however show the most difference in cost sensitivity for leisure trips, with the WTP for the latter being almost twice as high as for commute trips, whereas it is either insignificant or up to 30% higher for the other three classes. Interestingly, the '*Flex-sceptic individuals*' have the lowest ratio of cycling time to in-vehicle time, meaning that they perceive cycling time relatively less negatively than the other classes. In combination with other parameter estimates however (particularly the ASCs), this lower ratio of travel time does not contribute to a higher likelihood of the '*Flex-sceptic individuals*' to choose the bicycle.

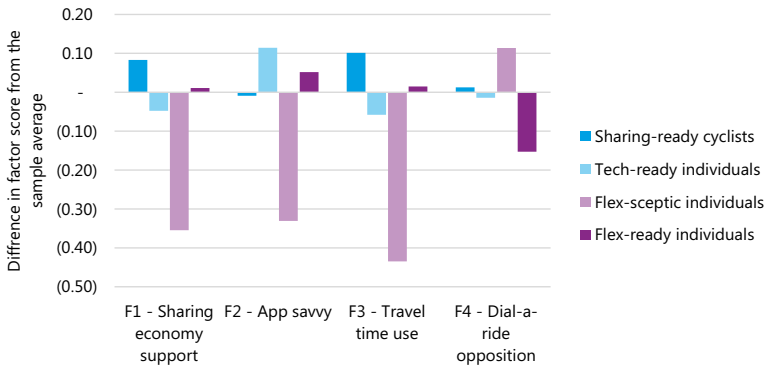

**Fig. 5** The differences between the sample average and the average of each of the four classes

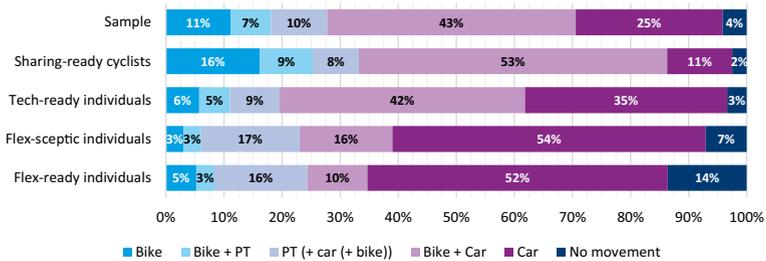

**Fig. 6** Weekly travel pattern of modes being used at least once per day, for the four latent classes

**Table 10** Socio-demographic characteristics of the sample and the four distinct latent classes

|  | Sample | Sharing-ready cyclists | Tech-ready individuals | Flex-sceptic individuals | Flex-ready individuals |
|---|---|---|---|---|---|
| *Gender* | | | | | |
| Female | 52% | 51% | 52% | 43% | 66% |
| Male | 48% | 49% | 48% | 57% | 34% |
| *Age* | | | | | |
| 18–34 | 21% | 22% | 23% | 16% | 16% |
| 35–49 | 19% | 18% | 23% | 23% | 14% |
| 50–64 | 30% | 31% | 27% | 23% | 39% |
| 65+ | 29% | 29% | 26% | 38% | 30% |
| *Education* | | | | | |
| Low | 30% | 29% | 29% | 36% | 37% |
| Middle | 41% | 38% | 44% | 47% | 45% |
| High | 28% | 33% | 27% | 17% | 18% |
| *Household income* | | | | | |
| Below average | 24% | 25% | 22% | 26% | 25% |
| Average | 23% | 24% | 18% | 29% | 22% |
| Above average | 40% | 40% | 45% | 25% | 35% |
| *Employment status* | | | | | |
| Employed | 50% | 50% | 52% | 44% | 47% |
| Student | 5% | 6% | 6% | 3% | 4% |
| Retired | 26% | 27% | 23% | 30% | 26% |
| Other non-employed | 19% | 18% | 19% | 23% | 24% |
| *Urbanisation level* | | | | | |
| Very highly urban | 23% | 25% | 20% | 23% | 24% |
| Highly urban | 32% | 31% | 33% | 34% | 29% |
| Moderately urban | 16% | 16% | 16% | 13% | 21% |
| Low urban | 20% | 20% | 20% | 18% | 22% |
| Not urban | 8% | 8% | 10% | 12% | 4% |
| *Household size* | | | | | |
| 1 | 22% | 23% | 22% | 22% | 23% |
| 2 | 36% | 38% | 32% | 34% | 38% |
| 3+ | 42% | 40% | 46% | 44% | 39% |
| *Household car ownership* | | | | | |
| Average | 1.14 | 1.07 | 1.27 | 1.11 | 1.19 |
| 0 | 17% | 20% | 11% | 17% | 18% |
| 1 | 56% | 56% | 56% | 59% | 55% |
| 2 | 23% | 22% | 28% | 20% | 19% |
| 3+ | 4% | 2% | 5% | 4% | 9% |

## Sharing-ready cyclists

This class is the most enthusiastic about cycling, strongly preferring it to all other modes. This class is not particularly sensitive to the in-vehicle time of motorised modes, but what makes the bike so attractive compared to any of the motorised modes is the price and the absence of walking and waiting times. The trip purpose does not play a role on their cost-sensitivity. Whether a Flex service is shared or private is also not relevant for their decision-making process.

This class has an above average positive opinion of the sharing economy and its merits, and they also see benefits in productively spending their travel time, even if it is not the shortest possible travel time. Their views on sharing and pre-booking a ride (F4) and on the use of apps are very much aligned with the sample. These results indicate that they seem to be (at least) partially ready to adopt Flex services.

'*Sharing-ready cyclists* 'are by far the most frequent bike and E-bike users, using them almost twice as often as any of the other classes, whereas their car use is the lowest among the classes (Fig. 6). They have the highest bike ownership, with 67% owning a bicycle and 34% an E-bike. They tend to be slightly younger, highest educated and most affluent. They are slightly more often found in highly urbanised areas, living without children (Table 10).

### Tech-ready individuals

Based on the SP experiment, the bicycle seems to be the most preferred mode for them, but less so than for the former class. '*Tech-ready individuals*' are more time-sensitive for in-vehicle time and are therefore prepared to pay more for a shorter travel time. For motorised modes, they have a willingness-to-pay ratio of 11.15 €/h. The ratio of cycling time to motorised mode in-vehicle time is also the highest of any class, meaning they perceive cycling time most negatively, making motorised modes comparatively more attractive. They are not as time-sensitive when it comes to out-of-vehicle time (walking and waiting), meaning they are willing to trade-off in-vehicle and out-of-vehicle time. '*Tech-ready individuals*' do not like sharing and are willing to pay €1.08 more to avoid it. They are also more cost-sensitive when making leisure trips, compared to commute trips, with a leisure trip WtP of 8.49 €/h.

This class is considered tech-ready, as they score highly on app-related statements, but low with respect to other factors. Interestingly, they have an above average familiarity with sharing economy services, being either most or second most familiar on all six examples, particularly in the frequency of using food delivery services.

'*Tech-ready individuals*' are frequent car users, with 86% using it on a weekly basis, with 53% using it daily. They also have the highest household car ownership, at 1.27 vehicles per household and only 11.5% of households have no car (compared to the 17% average). They have an above average bike ownership and are the second most frequent cyclists. Members of this class have a middle-to-high level of education (slightly above average) and have an above average household income. They are the youngest class, live more often in suburban areas and have the largest average household size, mostly living as a couple with children or a single parent with children.

### Flex-sceptic individuals

This class has a very strong preference for using the car, scoring well above all other modes. '*Flex-sceptic individuals*' are willing to spend most to reduce their travel time out of any of the classes, with a WtP of over 15€/h, whereas walking and waiting do not seem to play a role in their decision making (all three parameters insignificant). Sharing a Flex trip also does not appear to play a role in their mode choice. Their high time-sensitivity is only observed for commute trips however, as their WtP for leisure trips is only 7.96 €/h.

'*Flex-sceptic individuals*' have the most distinctive profile of the classes. They are strongly opinionated and quite strongly negative on all four factors resulting from the EFA (note that F4 is reversed, so an above average score indicates stronger opposition than the sample average).

Similar to their preferences in the survey, '*Flex-sceptic individuals*' are frequent car users, but with an average car ownership rate. They are less enthusiastic cyclists than the average and have a low bike ownership, with only 43% having a bike (low for the Dutch context), compared to the average of 62%. This class is on average the lowest educated and has the lowest income. They also live in less urbanised areas. The latter likely contributes to their above average preference for car, as other modes are comparatively less attractive in such a context. Although a below average income and high car preference may be counterintuitive, their car ownership rate is average for the sample, meaning they seem to be pragmatic in their car ownership. '*Flex-sceptic individuals*' are more likely to live alone, rather than with a partner (either with or without a child). The members of this class have an above average likelihood to be pensioners: 30%, compared to the sample average of 26%.

### Flex-ready individuals

Members of this class seem to have an overall minor preference for car over the other modes and also a clear, though less significant (p=0.092) preference for Flex over PT or bike. They are also the most cost-sensitive of the four classes. This makes '*Flex-ready individuals*' the most multimodal of all the classes: selecting a motorised mode when it is cheap and opting for bike when the others are too expensive. They do not like to walk, having the highest ratio of walking time to in-vehicle time in the sample at 2.5-times higher. Sharing Flex is barely significant for this class (p=0.10), with respondents willing to pay only as much as €0.53 more for a private ride, again highlighting their high cost sensitivity.

'*Flex-ready individuals*' are the strongest supporters of dial-a-ride-style services, as they do not seem to be too bothered by sitting close to strangers or having to pre-book a ride. They also seem to be fairly comfortable using apps, and score slightly above average in their favourable opinions towards the sharing economy and the use of travel time.

'*Flex-ready individuals*', like the '*Flex-sceptic individuals*', have an above average use of the car and below average use of bike. They have the lowest bike ownership of any class, at only 39%. Interestingly, they are also the most likely to not travel regularly on a weekly basis. This class has a both a slightly below average income and a slightly below average level of education. Given that many class members are young however, this may be because they have not yet completed their education. While the average age of this class is above the sample average, it has an above average share of both younger (below 25) and older individuals (above 60). They live in more urbanised areas in small households and more often live either alone or as a couple (without children).

## Discussion

Sharing an on-demand trip is often associated with two main differences compared to taking a private trip: physically sharing the vehicle with other passengers (close presence of strangers) and potential additional travel time. To compensate for this, a discount (financial incentive) is offered for sharing, but the discount necessary to attract different user groups varies based on their preferences. In Fig. 7, we show the discount needed for shared and private trips to be equally attractive (have the same utility) for the four user groups, given additional travel time. Although the variation of in-vehicle time due to detours was not studied, preliminary research in the field reports that additional travel time has an equal or lower VOT than the promised travel time at the start of the trip (Alonso-González et al. 2020c), meaning that our outcomes (Fig. 7) are more likely overestimating the necessary discount rather than underestimating it.

For larger travel time differences, two classes with a lower VOT and two classes with the higher VOT can be easily distinguished. Attracting '*Tech-ready individuals*' and '*Flex-sceptic individuals*' to a pooled alternative will therefore require a substantially larger discount than for '*Sharing-ready cyclists*' and '*Flex-ready individuals*'. This is also in-line with their attitudes towards sharing and sitting close to strangers, where '*Sharing-ready cyclists*' and '*Flex-ready individuals*' are more likely to not mind the proximity of strangers.

From Fig. 7, we would assume that '*Sharing-ready cyclists*' and '*Flex-ready individuals*' should be the target of Flex services, but that only compares shared and private Flex services, without yet positioning them in the wider mobility context. To that end, we perform a sensitivity analysis, mimicking a typical urban trip (Fig. 8), with a potential market share of a private or shared Flex service, based on a varied Flex trip cost. The findings are presented in Fig. 9.

'*Flex-ready individuals*' show the largest potential for using on-demand services. For them, the main competing modes are public transport and car, but as they are highly cost sensitive, Flex is a viable alternative as long as it is sufficiently cheap. '*Sharing-ready cyclists*' show almost no potential to use Flex in an urban environment. As their name suggests, cycling is their mode of choice and for short urban trips, they are unlikely to choose any other mode. Cycling only starts being less viable when it takes more than 30 min. The same can be said for '*Flex-sceptic individuals*', except in their case, the main competitor is the car. Given their strong preference for the car, it would have to be very strongly penalised (high fuel cost, parking cost, limited access,…) for any of the other modes to be a viable alternative. Because they also have a very negative perception of sharing, they may be the least likely to adopt a Flex service. '*Tech-ready individuals*' however, while not very open to sharing, are potential Flex users. Time and privacy are quite important to them so Flex needs to be fast, private and preferably also door-to-door. This is a premium that they are willing to pay for. Sharing on its own would not be a major issue, but since it would likely entail a longer travel time, it is out of the question for them.

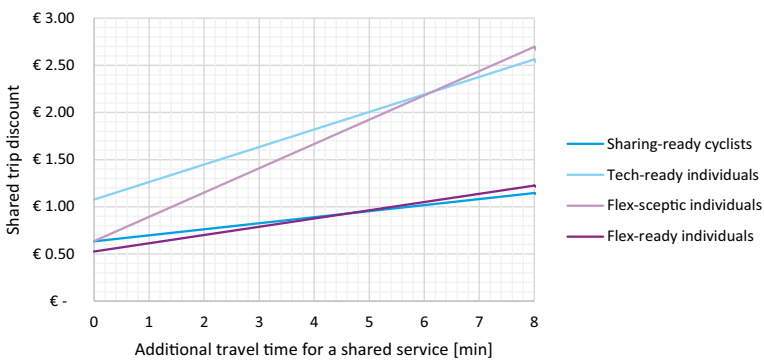

**Fig. 7** Discount required for a shared Flex ride to be equally attractive as a private ride

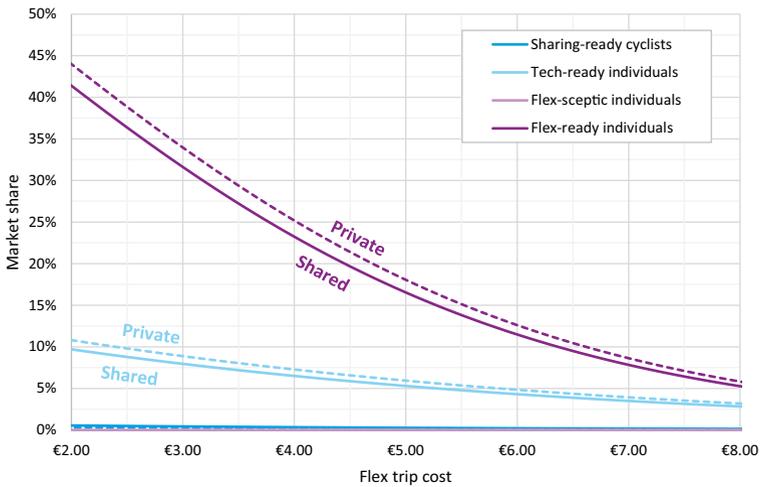

**Fig. 8** Example choice situation

**Fig. 9** Market share of Flex (private or shared, compared to a bicycle, car and PT alternative) among different classes when Flex trip cost is varied ceteris paribus

## Conclusion

This study analyses the role of on-demand transport services, both private and pooled, and their adoption potential within the context of competition for urban trips alongside alternative modes. A survey was carried out in the Netherlands, yielding 1,063 valuable responses. To the best of our knowledge, this is the first study to compare the expected use of on-demand services to car, public transport and bicycle transport and allow for the monetary evaluation thereof. Our survey took place in the Netherlands, where the cycling conditions and culture differ greatly from most previous studies which were set in North America, where cycling plays a much smaller role and where cars and (to a lesser extent) public transport carry the majority of urban travellers.

In a highly cycling-oriented environment, Flex services do not seem to offer a highly attractive alternative for (short) urban trips. In line with literature (Choudhury et al. 2018; Hyland et al. 2018; Liu et al. 2018; Yan et al. 2019), on-demand services fall behind cycling, the car and public transport, both in the overall mode perception as well as in the perception of cost. While cycling is not attractive over longer distances, most cities in the Netherlands and the wider region (Central and Western Europe) are relatively small and thus urban trips are usually not longer than five kilometres (de Graaf 2015). The rise of e-bikes is also extending the range for a journey by (e-)bike. While not of interest for commuting, Flex services are used more commonly around the world for leisure trips, mostly in the evenings and during the night (King et al. 2020; Mohamed et al. 2019; Young and Farber 2019). Our findings also support this, as the attractiveness of Flex—when compared to other modes—is found to be higher for leisure trips. Respondents are found to be more cost-sensitive when making leisure trips by car or PT, but not when choosing Flex.

To better understand market segmentation and how perceptions vary with respect to on-demand mobility, a latent class choice model is used in which four distinct market segments are uncovered. While being the smallest class (at 9% of the sample), the "*Flex-ready individuals*" have by far the highest propensity to choose a Flex service for an urban trip. With a relatively high cost-sensitivity and refraining from cycling, they are well positioned to adopt on-demand services if / when they become more commonplace. They are quite open to sharing a Flex ride and would only use a private service if it costs only a fraction more. The largest class of the sample (55%), the "*Sharing ready cyclists*" are also very open to sharing, exhibiting similar attitudes and behaviour with respect to sharing, but do not present a high potential for using Flex. As their name implies, they are avid cyclists, and for short urban trips there is virtually no alternative for them. A higher adoption potential can be observed among the "*Tech-ready individuals*" (27%), for whom time is of the essence. If Flex can provide them with a comparatively fast service, they are more than happy to use it. They are more averse to sharing, but if they do not lose much time, they would still consider it if they receive a significant discount for sharing. For the final class, the "*Flex-sceptic individuals*" (9%), on-demand services do not seem like a viable alternative. They display a dislike towards sharing, technology and use of public transport, while at the same time highly prioritising their car. Given these results, public transport—and to a lesser degree cars—will likely suffer the biggest impact of an increased presence of Flex in cities. Cycling will likely not be impacted greatly, as the most frequent cyclists do not perceive Flex as a viable alternative for them.

Recently, a few other studies identified latent clusters in the Dutch population, with respect to MaaS, shared, autonomous and on-demand mobility (Alonso-González et al. 2020a, b, c; Winter et al. 2020). Certain parallels can be drawn between their findings and ours. About half of the population seems to fall into a mostly-ready category with respect to MaaS and Flex, being both ready at the technology level, as well as the sharing and mobility aspect. They tend to cycle a lot, use the car below the average, are younger and higher educated. Around a quarter of the population is found to be so called 'technologi-cal car lovers', a group that is technologically advanced and neutral on sharing. Their main characteristic is they are very time-sensitive and are prepared to pay a lot. Two smaller classes (10–20% each) round off the population, with very opposing views. One class are confident in making multi-modal trip and for the most part ready to adopt MaaS. They live in urban areas, have a lower car ownership and are fairly cost-sensitive. On the other side is the class most negative about most things regarding technology and innovation, sticking to the privacy of their car. They tend to be middle-aged or older, predominantly male and middle educated.

A limitation of this study is the hypothetical bias that comes with using stated prefer-ence methods. Although some studies found limited bias, most conclude that respondents display a higher willingness-to-pay in the hypothetical setting of an SP survey (Loomis 2011; Murphy et al. 2005) and that actual behaviour may differ. The transferability of the results to other contexts is also limited. While cycling plays a dominant role for most short trips in the Netherlands, the attitude towards public transport may be lower in rural areas due to its lower quality and the perception of car more positive. While the former could make Flex comparatively more attractive to users, the latter could have a negative impact. Although great care was taken when selecting the attribute levels, to make sure a sufficiently large range of possible values is captured, they may still influence the model outcomes. Specifically with respect to the travel times of car, public transport and on-demand services, the range, although broad for the analysed context of an urban trip, does not vary greatly (between 8 and 16 min) and thus may have influenced the model estima-tion, wherein the parameter for the travel time of those modes was insignificant. The Dutch (north-west European) environment, with its high quality public transport and extensive cycling infrastructure also means that the results cannot be directly translated to other geo-graphic areas with minimal cycling and/or public transport use. With respect to the mar-ket segmentation, the "*sharing-ready cyclist*" group is likely (much) smaller and the three remaining classes larger, the extent to which varying depending on local cycling, traffic and public transport conditions.

In contexts where the use on-demand mobility services rises, policymakers need to make sure that active modes stay attractive and especially that public transportation does not start suffering from a decline in ridership due to travellers shifting modes. Munici-palities should cooperate with Flex service providers, to stimulate Flex as a complementary service to existing public transport services, by acting as a feeder, providing services in areas and at times of day with poor public transport services. To encourage pooling as much as possible, financial incentives (discounts) should also be offered to travellers for pooling their trips.

In future research, a mixed logit model can provide an alternative interpretation of the heterogeneity of users and their perception of Flex services. Furthermore, a latent class model with a class membership function that incorporates attitudinal and socio-demographic data or directly interacting this data with parameter estimates in the model can help in further identifying potential user groups for Flex services. Future research should also evaluate the potential of Flex for other trip types. An often cited case for (autonomous) on-demand services is for access/egress to/from train stations or other high capacity public transport (Chen and Nie 2017; Clements and Kockelman 2017; Hall et al. 2018; Ma 2017; OECD 2015; Tirachini and del Río 2019) and how such a service can impact access mode and station choice—potentially increasing the catchment area of stations. Another potential use of shared Flex services could be on medium-distance (up to 100 km) inter-city trips. Research should also consider how Flex services can aid the attractiveness of MaaS bundles, to complement public transport services by offering first/last mile access and accessibility at times of the day when public transport service are limited or non-existent, offering MaaS users a greater variety of alternatives. Finally, if / when Flex becomes more commonplace, its impact on road congestion, public transport ridership, vehicle ownership etc. needs to be evaluated, along with potential policies on how to increase its level-of-service while securing its affordability and financial viability and minimizing the negative externalities of transportation.

**Table 11** Model estimation results of the ASP model

| | Parameter estimate | Robust t-stat | Significance |
| --- | --- | --- | --- |
| Constant [bike] | 0 [ fixed] | | |
| Constant [car] | − 1.498 | − 10.37 | *** |
| Constant [Flex] | − 2.044 | − 9.68 | *** |
| Constant [PT] | − 2.082 | − 12.03 | *** |
| Cost [car] | − 0.094 | − 10.53 | *** |
| Cost [Flex] | − 0.408 | − 16.44 | *** |
| Cost [PT] | − 0.221 | − 7.50 | *** |
| Leisure trip * cost [car] | − 0.024 | − 2.56 | ** |
| Leisure trip * cost [Flex] | 0.002 | 0.11 | |
| Leisure trip * cost [PT] | − 0.061 | − 2.03 | ** |
| In-vehicle time [bike] | − 0.070 | − 11.21 | *** |
| In-vehicle time [car] | − 0.010 | − 1.38 | |
| In-vehicle time [Flex] | − 0.016 | − 1.34 | |
| In-vehicle time [PT] | − 0.012 | − 1.29 | |
| Walking time [car] | − 0.035 | − 5.73 | *** |
| Walking time [Flex] | − 0.104 | − 6.35 | *** |
| Walking time [PT] | − 0.055 | − 5.77 | *** |
| Waiting time [Flex] | − 0.022 | − 1.73 | * |
| Waiting time [PT] | − 0.040 | − 4.27 | *** |
| Sharing [Flex] | − 0.198 | − 2.56 | ** |

$***p \leq 0.01$, $**p \leq 0.05$, $*p \leq 0.1$

**Table 12** Estimation results of the DCP model

| Mode | Attribute | Level | Value | Robust t-stat | Significance |
|---|---|---|---|---|---|
| Bike | Constant | | *0 (fixed)* | – | – |
| | In-vehicle time | 12 min | *0 (fixed)* | – | – |
| | | 16 min | − 0.2139 | − 4.19 | *** |
| | | 20 min | − 0.5556 | − 11.21 | *** |
| Public transport | Constant | | − 1.5765 | − 18.27 | *** |
| | Travel cost | € 0.5 | *0 (fixed)* | – | – |
| | | € 2 | − 0.4356 | − 4.84 | *** |
| | | € 3.5 | − 0.6308 | − 6.56 | *** |
| | Leisure trip (additional impact) | € 2 | − 0.0974 | − 0.87 | |
| | | € 3.5 | − 0.2097 | − 1.68 | * |
| | In-vehicle time | 8 min | 0.0437 | 0.59 | |
| | | 12 min | *0 (fixed)* | – | – |
| | | 16 min | − 0.0591 | − 0.77 | |
| | Walking time | 1 min | *0 (fixed)* | – | – |
| | | 5 min | − 0.2075 | − 2.85 | *** |
| | | 9 min | − 0.4361 | − 5.67 | *** |
| | Waiting time | 1 min | *0 (fixed)* | – | – |
| | | 5 min | − 0.1207 | − 1.64 | |
| | | 9 min | − 0.3201 | − 4.19 | *** |
| Car | Constant | | − 0.8482 | − 12.32 | *** |
| | Travel cost | € 1 | *0 (fixed)* | – | – |
| | | € 5 | − 0.4960 | − 6.90 | *** |
| | | € 9 | − 0.7328 | − 9.59 | *** |
| | Leisure trip (additional impact) | € 5 | − 0.1636 | − 1.83 | * |
| | | € 9 | − 0.1754 | − 1.83 | * |
| | In-vehicle time | 8 min | 0.0698 | 1.17 | |
| | | 12 min | *0 (fixed)* | – | – |
| | | 16 min | − 0.0175 | − 0.29 | |
| | Walking time | 0 min | *0 (fixed)* | – | – |
| | | 5 min | − 0.1657 | − 2.81 | *** |
| | | 10 min | − 0.3488 | − 5.74 | *** |
| Mobility-on-demand | Constant | | − 2.1523 | − 17.87 | *** |
| | Travel cost | € 2 | *0 (fixed)* | – | – |
| | | € 5 | − 1.5182 | − 10.79 | *** |
| | | € 8 | − 2.2061 | − 11.96 | *** |
| | Leisure trip (additional impact) | € 5 | 0.2414 | 1.38 | |
| | | € 8 | − 0.2404 | − 0.89 | |
| | In-vehicle time | 8 min | 0.1691 | 1.81 | * |
| | | 12 min | *0 (fixed)* | – | – |
| | | 16 min | − 0.0061 | − 0.06 | |
| | Walking time | 0 min | *0 (fixed)* | – | – |
| | | 3 min | − 0.3282 | − 3.35 | *** |
| | | 6 min | − 0.7155 | − 6.61 | *** |
| | Waiting time | 1 min | *0 (fixed)* | – | – |

**Table 12** (continued)

| Mode | Attribute | Level | Value | Robust t-stat | Significance |
|---|---|---|---|---|---|
| | | 5 min | − 0.3095 | − 3.02 | *** |
| | | 9 min | − 0.1991 | − 1.95 | * |
| | Sharing the ride | | − 0.1549 | − 1.82 | * |

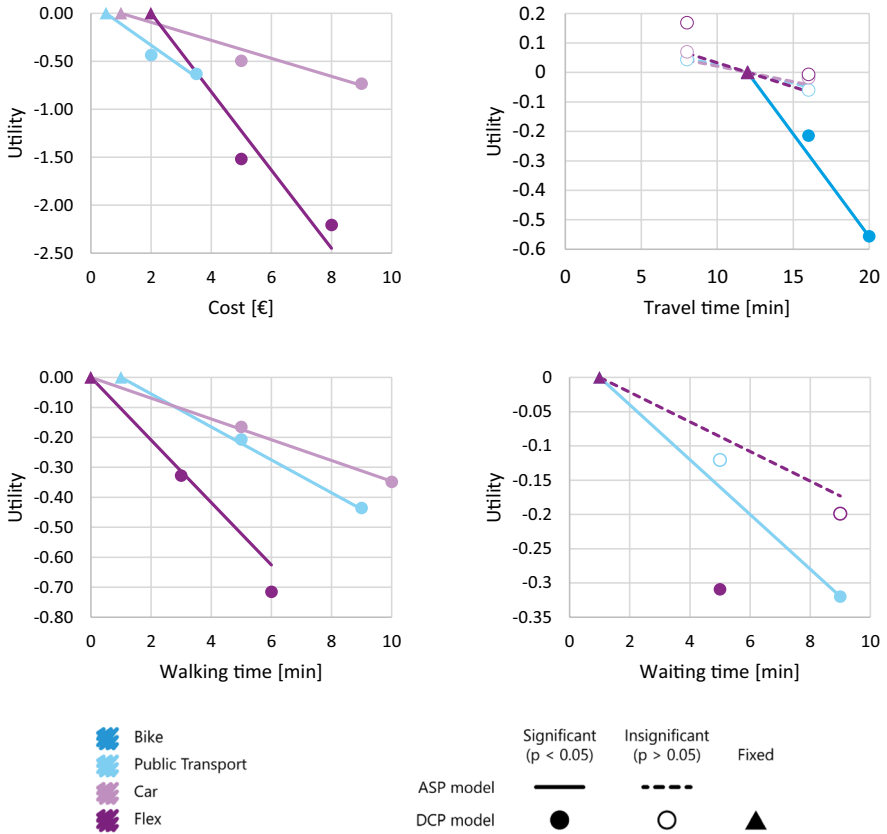

*for a clearer comparison of the parameters from the different model specification, the alternative-specific parameters are aligned to 0 at the reference level of the dummy-coded parameters*

**Fig. 10** Utility contributions of the linear alternative-specific parameters and the dummy parameters for the attributes of cost, travel time, walking time and waiting time

# Appendix 1

In this appendix, the outcomes of the ASP (Table 11) and DCP (Table 12) models are presented, along with a visualisation of the impact of the parameter estimates on the utility (Fig. 10). Most parameters seem to have a mostly linear impact on the utility of an alternative, which the ASP model can capture just as well as the dummy-coded model.

Adjusting the ASP model parameters to the reference level of the DCP model parameters, the estimated dummy parameters largely follow a linear pattern with only minor deviations from the alternative-specific parameters. With respect to cost, a slight reduction of marginal utility contribution can be seen (each additional euro has a lower impact on utility than the previous one), whereas with all three time-related parameters, a slightly increasing marginal utility contribution (each additional minute of travel/walking/waiting contributes more disutility than the previous one) can be observed. A notable exception here is the waiting time for Flex. As explained in Sect. 2, Flex waiting time and PT waiting time are expected to be perceived and experienced differently (and were therefore presented differently in the survey), and this is indeed confirmed by the model estimation results. What is most striking about the Flex waiting time is that a 5-min waiting time seems to have a more negative impact than a 9-min waiting time. As explained in Sect. 4.1, this could be indicating respondents perceiving waiting at home as hidden waiting time.

**Acknowledgements** This research was supported by the CriticalMaaS project (Grant Number 804469), which is financed by the European Research Council and Amsterdam Institute for Advanced Metropolitan Solutions.

**Author contribution** The authors confirm contribution to the paper as follows: Study conception and design: NG, OC, NO. Survey design and data gathering: NG, SHL, NO, OC. Analysis and interpretation of results: NG, NO, OC. Draft manuscript preparation: NG, NO, OC. Supervision and reviewing: OC, NO, SHL. Funding acquisition: OC. All authors reviewed the results and approved the final version of the manuscript.

## Declarations

**Conflict of interest** On behalf of all authors, the corresponding author states that there is no conflict of interest.

**Nejc Geržinič** is a PhD candidate at the Department of Transport & Planning, at Delft University of Technology. His main fields of interest are travel behaviour research and choice modelling on the topics of public transportation and sustainable mobility in the future.

**Niels van Oort** is an Assistant Professor at the Department of Transport & Planning, at Delft University of Technology. He is the co-director of the Smart Public Transport Lab. His main fields of expertise are public transport planning, dealing with the passenger perspective, service reliability and big data.

**Sascha Hoogendoorn-Lanser** is the director of the Mobility Innovations Centre Delft. Her expertise lie in travel behaviour research and stated preference data collection methods.

**Oded Cats** is an Associate Professor at the Department of Transport & Planning, at Delft University of Technology. He is the co-director of the Smart Public Transport Lab. His main research aim is to develop theories and models of multi-modal passenger transport networks by combining advancements from behavioral sciences, operations research and complex network theory.

**Serge Hoogendoorn** is a Professor and chair at the Department of Transport & Planning, at Delft University of Technology. His current research focuses on various aspects of Smart Urban Mobility.